\documentclass[journal]{IEEEtran}

\usepackage{myStyleIEEE}
\newtheorem{proposition}{Proposition}

\newtheorem{theorem}{Theorem}

\DeclareSIUnit \voltampere { VA } 
\DeclareSIUnit \var { var } 

\IEEEoverridecommandlockouts

\begin{document}

\title{Online Feedback Droop Scheduling in Distribution Grids for Frequency and Local Voltage Control}

\renewcommand{\theenumi}{\alph{enumi}}

\newcommand{\uros}[1]{\textcolor{magenta}{$\xrightarrow[]{\text{U}}$ #1}}
\newcommand{\petros}[1]{\textcolor{red}{$\xrightarrow[]{\text{P}}$ #1}}
\newcommand{\ognjen}[1]{\textcolor{pPurple}{$\xrightarrow[]{\text{O}}$ #1}}

\author{Ognjen~Stanojev,~\IEEEmembership{Student~Member,~IEEE,}
        Yi~Guo,~\IEEEmembership{Member,~IEEE,}
        Gabriela~Hug,~\IEEEmembership{Senior~Member,~IEEE}

\thanks{This research was supported by the Swiss National Science Foundation under NCCR Automation, grant agreement 51NF40\_180545. This work was partially supported by an ETH Z\"{u}rich Postdoctoral Fellowship}
\thanks{O. Stanojev, Y. Guo, and G. Hug are with Power Systems Laboratory at ETH Z\"{u}rich, Z\"{u}rich, 8092,  Switzerland, emails:\{guo, stanojev, hug\}@eeh.ee.ethz.ch. }
}

\maketitle
\IEEEpeerreviewmaketitle

\begin{abstract}
This paper presents a novel framework for collective control of Distributed Energy Resources (DERs) in active Distribution Networks (DNs). The proposed approach unifies the commonly employed local (i.e., decentralized) voltage and frequency droop control schemes into a transfer matrix relating frequency and voltage magnitude measurements to active and reactive power injection adjustments. Furthermore, the transfer matrices of individual DER units are adaptively tuned in real-time via slow communication links using a novel online gain scheduling approach to enable primary frequency support provision to the transmission system and ensure that the DN voltages are kept within the allowable limits. A global asymptomatic stability condition of the analyzed droop-controlled DN is analytically established. The considered gain scheduling problem is solved by leveraging an online primal-dual gradient-based method and a suitable linearized power flow model. Additional ancillary service providers can be trivially incorporated into the proposed framework in a plug-and-play fashion. Numerical simulations of the 37-bus IEEE test system and a realistic Swedish 533-bus DN confirm the validity and the scalability of the approach and demonstrate numerous advantages of the proposed scheme over the state-of-the-art.
\end{abstract}


\section{Introduction} \label{sec:intro}
Distribution grids are currently undergoing a period of radical changes governed by the proliferation of inverter-interfaced Distributed Energy Resources (DERs) and growing load demand in the form of electric vehicles. Deployment of such technologies has brought about new challenges in the operation of distribution networks, as high bidirectional power flows result in frequent overvoltage and undervoltage scenarios. The rapid fluctuations in bus voltages cannot be contained using slowly responding traditional voltage control resources such as load tap changers and capacitor banks \cite{Baran1989}, thus incentivizing the grid operators to untap the voltage control potential of the deployed DERs \cite{Turitsyn2011}. The operation of contemporary distribution grids is becoming even more difficult due to newly introduced requirements for the provision of regulation services to the transmission system \cite{Hatziargyriou2017}. To assist in this task, collections of heterogeneous DERs in distribution grids can be aggregated and controlled as a virtual power plant, providing frequency support through regulation of the active power exchange with the transmission system \cite{Emiliano2018}.


The aforementioned voltage control problem of active distribution grids has been extensively studied by both practitioners and academics for more than a decade. While early works have demonstrated the efficiency and benefits of centralized control schemes that solve a global optimal power flow problem \cite{Gabash2012}, most of the research has been directed towards the development of purely local (i.e., decentralized) control strategies \cite{Tonkoski2011,Jahangiri2013,Samadi2014,StavrosDataDriven2019,Farivar2013,HaoZhu2016,Xinyang2021} due to the absence of fast communication and monitoring infrastructure in distribution networks. The most common design of the local control law is the so-called droop control, where the DER reactive power injection is computed as a (piecewise) linear function of the measured voltage amplitude at the interface bus. Additionally, the droop relationship between the DER active power injection and the measured voltage amplitude can be employed to achieve superior controllability of the distribution grid voltages \cite{Tonkoski2011,KyriDroop2018}. While the droop schemes offer low-cost, scalable, and communication-free control, the two main challenges lie in ensuring the system stability and selecting the droop gains. 

Stability concerns for distribution grids with droop-based voltage control have first been raised in \cite{Jahangiri2013}, where high droop gains were noticed to lead to undesirable oscillatory behavior of bus voltages. In \cite{Farivar2013,HaoZhu2016,Xinyang2021}, analytical bounds on droop gains have been derived by reverse-engineering the distribution grid voltage regulation problem and analyzing the convergence of the obtained distributed algorithm. Input-to-state stability properties of droop control were considered in \cite{KyriDroop2018}, while \cite{Eggli2021} attained conditions for global asymptotic stability. However, the aforementioned approaches focus on discrete-time models and disregard the dynamics of inverter control loops. The stability of droop-controlled DERs with a realistic continuous-time inverter model was examined in \cite{Filip2015} by performing root locus analysis and experimental tests. 

A recommended parametrization of the voltage droop control for DERs is specified in the IEEE 1547 standard~\cite{IEEE1547}. However, the proposed tuning is network-agnostic and does not guarantee the satisfaction of voltage constraints \cite{KyriDroop2018}. Optimization-based offline droop tuning using network models was analyzed in \cite{Samadi2014,Calderaro2014}, whereas a data-driven approach was proposed in \cite{StavrosDataDriven2019}, approximating the optimal droop control parameters using support vector machine regression. However, the offline parameter selection approaches suffer high computational complexity and, as demonstrated in \cite{Saverio2019}, do not guarantee the desired regulation for all possible equilibrium points. Higher performance can be achieved by adaptively adjusting the droop gains via communication links on a regular basis to account for changes in weather conditions, load patterns, or the network topology. Such an approach was considered in \cite{KyriDroop2018}, where the coefficients of the proportional controllers are obtained by solving a robust optimization problem and updated online every 5-15 minutes. 

Apart from the previously discussed voltage control problem, there is a growing need to incorporate frequency control functionalities into DERs in distribution grids due to the decommissioning of many traditional frequency providers, i.e., synchronous generators. Contrary to the voltage control problem, the design of local frequency droop controllers for DERs in distribution grids has been less addressed in the literature thus far, as most recent works focus on centralized \cite{StavrosAS2020,OgnjenPowerTech2021,stanojev2022multiple} and distributed \cite{Tang2018} frequency control provision schemes. Furthermore, the decentralized frequency control principles of large synchronous machines cannot trivially be transferred and adopted for the control of dispersed DERs. The range of feasible operating points of distribution grids is limited by the static and dynamic properties of the network components and operational circumstances. Moreover, the availability of the resources varies substantially during the day and depends on numerous exogenous factors. Thus, there is no one-size-fits-all solution for the selection of the frequency droop gains for the individual DERs \cite{OgnjenPowerTech2021}. The gains need to be adaptively tuned based on the state of the distribution network and the availability of DER resources while ensuring that the distribution grid as a whole provides the required amount of frequency support to the transmission grid.

In this work, we propose a novel framework for local voltage and frequency control in distribution grids, where generalized $2\times2$ droop gain matrices are used for fast local control of DERs, and a supervisory controller is used to update the droop gains to account for the changing grid conditions. The contributions of this work can be listed as follows: 

\begin{itemize}
    \item Modeling and local control of the inverter-based DERs is enhanced compared to the state-of-art \cite{KyriDroop2018,Xinyang2021} by considering dynamics of the current controllers, i.e., the inner control loops, and by using the generalized droop gain matrix instead of the decoupled droop control. Furthermore, novel insights into the stability properties of droop-controlled distribution grids are analytically established by constructing a Lyapunov function that certifies the global asymptotic stability of the obtained closed-loop system. Stability conditions for selecting the droop gains are thus obtained and can be used in the control design.
    \item We formulate an optimal droop scheduling problem to adaptively tune the droop gain matrices of individual DERs. In this way, the desired regulation requirements can be satisfied in spite of the changing operational conditions. The divide and conquer strategy \cite{Leith2000SurveyOG} is used to decompose the nonlinear continuous-time control task into simpler linear control design problems for a set of operating points. The droop coefficients can be obtained for each operating point by solving a stochastic optimization problem with chance constraints. The optimization problem embeds the regulation requirements of keeping the bus voltages within allowable limits and providing sufficient active power to support the transmission system frequency while minimizing the regulation costs. The stochastic formulation is adopted to manage uncertainty pertaining to uncontrollable power injections.
    \item A novel way of solving the droop scheduling problem using the primal-dual gradient algorithm is proposed. This solution approach results in an online feedback controller that continuously steers the system towards the solutions of the above-mentioned stochastic optimization problem by adaptively tuning the generalized droop gain matrix depending on the time-varying condition of the distribution grid. A sequence of linear controllers is thus used to drive the distribution network to an operating point where the regulation requirements are satisfied at the lowest cost. The online feedback-based solution approach has been selected as it possess high computational efficiency (at each time step only matrix multiplications have to be performed to obtain the solution), scalability and plug-and-play features. Both centralized and distributed implementations of the proposed algorithm are viable.  In contrast to \cite{KyriDroop2018} where a centralized quadratic optimization problem is solved to update the gains, the proposed approach offers simpler implementation and is computationally more efficient. The proposed method demonstrates computational advantages over previous hierarchical control methods \cite{Zhang2020,Zhang2022} as well. Compared to online schemes which dispatch the power setpoints to DERs to track solutions of an Optimal Power Flow (OPF) problem \cite{Emiliano2018,EmilianoOPFPursuit}, the proposed method capitalizes both on global coordination and fast local control, thus improving the regulation performance.
\end{itemize}


The rest of the paper is structured as follows. Section~\ref{sec:preliminaries} introduces the considered distribution grid and DER models, and establishes the stability conditions. The online droop scheduling strategy is presented in Sec.~\ref{sec:OFDS}, where the adaptive divide and conquer strategy is first introduced. Then, a stochastic optimization problem is formulated to find the optimal droop gains at each time step, and lastly, the proposed optimization problem is solved using a primal-dual gradient algorithm. Finally, performance of the proposed control framework is tested by performing multiple case studies in Sec.~\ref{sec:res}.

\textit{Notation}. We denote the sets of real and natural numbers by $\mathbb{R}$ and $\mathbb{N}$, and define $\R_{\geq a}\coloneqq\{x\in\R\mid x\geq a\}$. Given a matrix $A$, $A^\top$ denotes its transpose. For column vectors $x\in\R^n$ and $y\in\R^m$ we use $(x,y)\coloneqq [x^\top,y^\top]^\top\in\R^{n+m}$ to denote a stacked vector. The largest/smallest eigenvalue of $A$ is represented by $\lambda_{\mathrm{max}}(A)/\lambda_{\mathrm{min}}(A)$. We write $A \preceq 0$ ($A \prec 0$) to denote that $A$ is negative semidefinite (definite). Finally, $I$ denotes the identity matrix of appropriate dimension.

\section{Preliminaries on Distribution System Modelling, Control and Stability} \label{sec:preliminaries}

\subsection{Network and Device Models}
This work considers a radial balanced distribution network represented by a connected graph $\mathcal{G}=(\mathcal{N},\mathcal{E})$, with $\mathcal{N} \coloneqq \{0,1,\dots,n\}$ denoting the set of network nodes including the substation node $0$, and $\mathcal{E} \subseteq \mathcal{N}\times\mathcal{N}$ representing the set of $n\in\mathbb{N}$ network branches. A single Point of Common Coupling (PCC) to the transmission grid at node 0 is assumed. The distribution network hosts a number of DERs and loads, where $\mathcal{C}\subseteq\mathcal{N}$ denotes the set of nodes with controllable converter-interfaced DERs of cardinality $n_c\coloneqq|\mathcal{C}|$. 
For every bus $i\in\mathcal{N}$, let $v_i\in\R_{\geq 0}$ denote the voltage magnitude, and $p_{c_i}\in\R$ and $q_{c_i}\in\R$ represent the active and reactive power injections of controllable DERs. The active and reactive power injections of uncontrollable devices, designated by $p_{d_i}\in\R$ and $q_{d_i}\in\R$, are modelled as random variables to capture the random load fluctuations and uncertain injections of renewable sources. For each branch $(i,j)\in\mathcal{E}$, let $r_{ij}\in\R_{\geq 0}$ and $x_{ij}\in\R_{\geq 0}$ denote its respective resistance and reactance values, $P_{ij}\in\R$ and $Q_{ij}\in\R$ denote the real and reactive power flow along the branch, and $i_{ij}\in\R_{\geq 0}$ represent the corresponding branch current magnitude. The distribution grid is modelled using the DistFlow equations \cite{Baran1989}, written recursively for every line $(i,j)\in\mathcal{E}$ as:
\begin{subequations} \label{eq:distFlow}
\begin{align}
    P_{ij}  &= \sum_{k\in\mathcal{N}_j} P_{jk} - p_{c_j} - p_{d_j} + r_{ij}i_{ij}^2, \label{eq:p_distFlow} \\
    Q_{ij}  &= \sum_{k\in\mathcal{N}_j} Q_{jk} - q_{c_j} - q_{d_j} + x_{ij}i_{ij}^2, \label{eq:q_distFlow} \\
    v_i^2 - v_j^2 &= 2(r_{ij}P_{ij}+x_{ij}Q_{ij}) - (r_{ij}^2+x_{ij}^2)i_{ij}^2, \label{eq:v_distFlow}
\end{align}
\end{subequations}
with the branch currents computed as $i_{ij}^2=(P_{ij}^2+Q_{ij}^2)/v_i^2$.


To enhance modeling of the controllable DERs, we consider the dynamics of the inverter inner current control loops. The response of inverter-based DERs can thus be described with the required level of accuracy by the first-order filter dynamics:
\begin{subequations}\label{eq:DER_model}
\begin{align}
    \tau_{p_i}\,\ddt{p}_{c_i} &= u_{p_i}  - p_{c_i}, \quad \forall i\in\mathcal{C}, \label{eq:p_DER} \\
    \tau_{q_i}\,\ddt{q}_{c_i} &= u_{q_i}  - q_{c_i}, \quad\, \forall i\in\mathcal{C}, \label{eq:q_DER}
\end{align}
\end{subequations}
where $p_{c_i}\in\R$ and $q_{c_i}\in\R$ denote the DER active and reactive power outputs, $\tau_{p_i}\in\R_{\geq0}$ and $\tau_{q_i}\in\R_{\geq0}$ are the related time constants, and $u_{p_i}\in\R$ and $u_{q_i}\in\R$ are the control inputs to the inner control loops. Experimental evidence on a two inverter test setup presented in \cite{Filip2015} confirms the model. 

\subsection{Local Control of DERs}
The principal objective of distribution grid controls is to maintain nodal voltages within prescribed limits. More precisely, given any system operating condition, the control goal is to steer the system voltage to reach the allowable range at the lowest cost. Secondarily, support to the transmission system should be provided by appropriately adjusting the power flow at the PCC in response to frequency deviations. The requirement of the Transmission System Operator (TSO) for the provision of primary frequency control is reflected in the active power adjustment that is to be applied to the power exchange $\Delta p_\mathrm{f}\in\R$ between the transmission and the distribution system in response to frequency $\omega\in\R_{\geq 0}$ variations around the nominal value $\omega^\star\in\R_{\geq 0}$:
\begin{equation} \label{eq:pfc_provision}
    \Delta p_\mathrm{f} = K^\mathrm{pf}_\mathrm{agg}(\omega-\omega^\star).
\end{equation}
The amount of regulating power to be provided is typically contracted through a market-based mechanism or directly prescribed by the TSO in the form of the gain $K^\mathrm{pf}_\mathrm{agg}\in\R_{\leq 0}$.

The DER units have two control resources that can be used to satisfy the control objectives, namely, adjustment of active and reactive power injections as shown in \eqref{eq:DER_model}. Traditionally, active power is used for frequency control and reactive power for voltage control. Nonetheless, with low X/R ratios in distribution grids, the converse relationships become viable as well \cite{Bevrani2013}. The inputs of inverter-based DERs with generalized droop-based frequency and local voltage controls can be described for each unit $i\in\mathcal{C}$ as follows:
\vspace{0.1cm}
\begin{equation}\label{eq:DER_inputs}
    \begin{bmatrix}
        u_{p_i} \\ u_{q_i}
    \end{bmatrix}
    = 
    \begin{bmatrix}
        p^\star_{c_i} \\ q^\star_{c_i}
    \end{bmatrix}
    +
    \underbrace{\begin{bmatrix}
        k_i^\mathrm{pv} & k_i^\mathrm{pf} \\ k_i^\mathrm{qv} & k_i^\mathrm{qf}
    \end{bmatrix}}_{\eqqcolon K_i}
    \begin{bmatrix}
        v_i-v^\star \\ \omega - \omega^\star
    \end{bmatrix}
\end{equation}
where $p^\star_{c_i}\in\R$ and $q^\star_{c_i}\in\R$ denote the active and reactive power output references, set via an economic dispatch problem. The local measurements of voltage magnitude and frequency are denoted by $v_i$ and $\omega$, respectively, and the nominal voltage of the grid is indicated by $v^\star\in\R_{\geq0}$. Droop gains pertaining to the adjustment of the active power output in response to voltage and frequency deviations for unit $i\in\mathcal{C}$ are denoted by $k_i^\mathrm{pv}\in\R$ and $k_i^\mathrm{pf}\in\R$, respectively. Similarly, droop gains pertaining to the reactive power output adjustment in response to voltage and frequency deviations are denoted by $k_i^\mathrm{qv}\in\R$ and $k_i^\mathrm{qf}\in\R$. Thus, a generalized $2\times 2$ droop gain matrix $K_i$ governing the primary response of DER $i\in\mathcal{C}$ to frequency and local voltage deviations at its interface node is obtained. Converter synchronization units, such as the Phase-Locked-Loop, are not modeled, and frequency $\omega$ is assumed to be precisely and concurrently measured at all nodes in the distribution grid. Such an approach has been shown to introduce only a marginal modeling error, as discussed in \cite{ORTEGA201837}.

The active and reactive power outputs of DER units need to comply with their hardware and operational constraints represented by the so-called capability curves, defined as sets of allowable setpoints $(p_{c_i},q_{c_i}) \in\mathcal{X}_{i}\subseteq\R^2$ for each unit $i\in\mathcal{C}$. For example, the operating region of the PhotoVoltaics (PVs) is defined by the minimum power factor and apparent power constraints, and the flexible load operation is limited by the constant power factor and minimum and maximum power output values. Note that the maximum available active power of PV and flexible load units changes with solar irradiation and load consumption. Thus, the capability sets of DERs are time-varying. The presented DER constraints can be enforced by implementing projections of $(u_{p_i}-p_{c_i},u_{q_i}-q_{c_i})$ onto $\mathcal{X}_i$ in the low-level inverter control logic for each unit $i\in\mathcal{C}$.

\subsection{Linear System Model}
To facilitate the stability analysis and the control algorithm design, we leverage a linear approximation of \eqref{eq:distFlow} given by the following model:
\begin{equation} \label{eq:LinDistFlow}
    v = R(p_c + p_d) + X(q_c+q_d)+v_0,
\end{equation}
where $v\coloneqq(v_1,\dots,v_{n})$ collects voltage magnitudes, and $p_c\coloneqq(p_{c_1},\dots,p_{c_n})$ and $q_c\coloneqq(q_{c_1},\dots,q_{c_n})$ are the vectors of active and reactive power bus injections of controllable DER units. Similary, $p_d$ and $q_d$ are obtained by stacking the uncontrollable injections. The voltage-active power sensitivity matrix $R\in\R^{n\times n}$, voltage-reactive power sensitivity matrix $X\in\R^{n\times n}$, and linearization point $v_0\in\R^n$ can be obtained using any power flow linearization method \cite{Christakou2013,Bolognani2016,HaoZhu2016}.
\begin{proposition} \label{prop:R_X_PDF}
The active and reactive power sensitivity matrices $R$ and $X$ are positive definite \cite{HaoZhu2016,Farivar2013}.
\end{proposition}
The small-signal counterpart of the closed-loop system defined in \eqref{eq:distFlow}-\eqref{eq:DER_model} and \eqref{eq:DER_inputs}, linearized around the nominal operating point using \eqref{eq:LinDistFlow}, is formulated in a compact matrix form as follows:
\begin{subequations} \label{eq:lin_sys}
\begin{align}
    \Delta v &= R\Delta p_c + X\Delta q_c + R\Delta p_d + X\Delta q_d,\label{eq:lin_sys_1} \\
    T_p\Delta\dot{p}_c &= - \Delta p_c + K^\mathrm{pv}\Delta v + K^\mathrm{pf}\Delta\omega,\label{eq:lin_sys_2} \\ 
    T_q\Delta\dot{q}_c &= - \Delta q_c + K^\mathrm{qv}\Delta v + K^\mathrm{qf}\Delta\omega,\label{eq:lin_sys_3}
\end{align}
\end{subequations}
where $\Delta$ denotes small deviations around the equilibirum. Time constants related to active and reactive power dynamics are collected in diagonal matrices $T_p\coloneqq\diag(\tau_{p_1},\dots,\tau_{p_n})$ and $T_q\coloneqq\diag(\tau_{q_1},\dots,\tau_{q_n})$, while entries of diagonal matrices $K^\mathrm{pv}\coloneqq\diag(k_1^\mathrm{pv},\dots,k_n^\mathrm{pv})$ and $K^\mathrm{qv}\coloneqq\diag(k_1^\mathrm{qv},\dots,k_n^\mathrm{qv})$ are populated with the respective voltage control-related droop gains. Vectors $K^\mathrm{pf}\coloneqq(k_1^\mathrm{pf},\dots,k_n^\mathrm{pf})$ and $K^\mathrm{qf}\coloneqq(k_1^\mathrm{qf},\dots,k_n^\mathrm{qf})$ collect the appropriate frequency control-related droop gains.
\subsection{Stability Analysis}
The frequency stability of droop-controlled transmission systems has been extensively studied \cite{Weitenberg2019,Cui2023}. Results from \cite{Cui2023} indicate that the local exponential frequency stability is achieved if droop coefficients of all generating units are selected to be non-positive. Thus, we require that $K^\mathrm{pf}_\mathrm{agg}\leq0$ and consequently $k_i^\mathrm{pf}\leq0,k_i^\mathrm{qf}\leq0,\forall i\in\mathcal{C}$ based on the linear model in \eqref{eq:lin_sys}.
The voltage stability properties are analyzed based on \eqref{eq:lin_sys} by identifying sufficient conditions for asymptotic stability. Injections pertaining to uncontrollable devices are random variables presented to the system as additive noise and hence, do not influence the system stability. In addition, since the frequency stability conditions have been established, the frequency control droop gains can be omitted from the voltage stability analysis. Therefore, by neglecting these terms and substituting \eqref{eq:lin_sys_1} into \eqref{eq:lin_sys_2} and \eqref{eq:lin_sys_3}, we obtain the following linear system representation:
\begin{equation} \label{eq:closed_loop_ssm}
    T\Delta\dot{x} = (K_\mathrm{v}G-I)\Delta x,
\end{equation}
with $\Delta x \coloneqq (\Delta p_c, \Delta q_c)$ and the model matrices defined as
\begin{equation*}
T =
    \begin{bmatrix}
     T_p & 0 \\
     0 & T_q
    \end{bmatrix},\,
K_\textrm{v} =
    \begin{bmatrix}
     K^\mathrm{pv} & K^\mathrm{pv} \\
     K^\mathrm{qv} & K^\mathrm{qv}
    \end{bmatrix},\,
G =
    \begin{bmatrix}
     R & 0 \\
     0 & X
    \end{bmatrix}.
\end{equation*}
\begin{theorem}\label{theorem:1}The closed-loop system \eqref{eq:closed_loop_ssm} is asymptotically stable if one of the following conditions, i.e., \eqref{eq:stability_1} or \eqref{eq:stability_2}, holds for all $i\in\mathcal{N}$:
\begin{equation}\label{eq:stability_1}
\begin{aligned}
&\left(\tau_{p_i}^{-1}k^{\mathrm{pv}}_i - \tau_{q_i}^{-1}k^{\mathrm{qv}}_i\right)^2 + 4\gamma \left(\tau_{p_i}^{-1}k^{\mathrm{pv}}_i - \tau_{q_i}^{-1}k^{\mathrm{qv}}_i\right) - 4\gamma^2 < 0,\\
&~~\tau_{p_i}^{-1}k^{\mathrm{pv}}_i - \gamma < 0,
\end{aligned}
\end{equation}
\begin{equation}\label{eq:stability_2}
\begin{aligned}
&\left(\tau_{p_i}^{-1}k^{\mathrm{pv}}_i - \tau_{q_i}^{-1}k^{\mathrm{qv}}_i\right)^2 + 4\gamma \left(\tau_{p_i}^{-1}k^{\mathrm{pv}}_i - \tau_{q_i}^{-1}k^{\mathrm{qv}}_i\right) - 4\gamma^2 < 0,\\
&~~\tau_{q_i}^{-1}k^{\mathrm{qv}}_i - \gamma < 0,
\end{aligned}
\end{equation}
where $\gamma = \lambda_{\mathrm{min}}((GT)^{-1})$.
\end{theorem}
 
\begin{proof}
    See Appendix.
\end{proof}

Inequalities \eqref{eq:stability_1} and \eqref{eq:stability_2} characterize the stability conditions of droop-controlled distribution grids, with the dynamics approximated by \eqref{eq:closed_loop_ssm}. It can be seen that the droop gains pertaining to active and reactive power are tightly interdependent. Furthermore, the network topology and parameters, appearing in the conditions via the sensitivity matrix $G$, also play an essential role. Notice that the conditions for each DER $i\in\mathcal{C}$ restrict solely the local gain pairs $(k_i^\mathrm{pv},k_i^\mathrm{qv})$ and depend on parameters $(\tau_{p_i},\tau_{q_i})$ and $\gamma$. Therefore, only the parameter $\gamma$ is common to the stability conditions of all DERs. This fact enables a decentralized implementation of the stability conditions. In the rest of this paper, we will leverage the decentralized property of this condition and encode it as the feasibility region in the optimal droop scheduling algorithm.

\section{Optimal Feedback Droop Scheduling: An Online Approach} \label{sec:OFDS}
This section discusses the main problem of continuously finding the optimal droop slopes to ensure the satisfaction of local voltage and primary frequency control provision constraints during the real-time operation of distribution grids. The central concept of droop scheduling is presented in Sec.~\ref{subsec:droop_scheduling}, followed by the proposed solution approach via the divide and conquer strategy in Sec.~\ref{subsec:divide_and_conquer}. Online implementation and the scheduling algorithm are presented in Sec.~\ref{subsec:OFO}.

\subsection{Towards Optimal Control via Droop Scheduling}\label{subsec:droop_scheduling}
Let $x\coloneqq(p_c,q_c)$ denote the vector of differential state variables, $d\coloneqq(p_d,q_d)$ designate the vector of uncontrollable (stochastic) injections, $r\coloneqq(p_c^\star,q_c^\star)$ denote the vector of active and reactive power DER references, and $y\coloneqq(v,P,Q,\Delta p_\mathrm{f})$ represent the vector of algebraic states, with vectors $P\in\R^n$ and $Q\in\R^n$ collecting active and reactive power flows of all branches. Then, the nonlinear Differential-Algebraic Equation (DAE) model of the distribution system can be written as:
\begin{subequations} \label{eq:sys_dae}
\begin{align}
    \dot{x} &= f(x,u),  \label{eq:diff_equation_general} \\
    0 &= g(x,y,d,\omega), \label{eq:alg_equation_general}
\end{align}
\end{subequations}
where $u\coloneqq(u_{p_1},\dots,u_{p_n},u_{q_1},\dots,u_{q_n})$ is the vector of DER inputs, $f(\cdot)$ is the state evolution function defined by \eqref{eq:DER_model}, and $g(\cdot)$ represents the nonlinear power flow model in \eqref{eq:distFlow} together with the frequency control requirement \eqref{eq:pfc_provision}. In summary, the DAEs in \eqref{eq:sys_dae} define the uncontrolled distribution system dynamics. The loop is closed by establishing droop (i.e., proportional) control on local voltage and frequency deviations \eqref{eq:DER_inputs}. Nevertheless, as indicated by \cite{Saverio2019,KyriDroop2018}, a single set of droop gains does not guarantee the desired regulation and therefore, an adaptive adjustment of the droop gains during real-time operation is necessary. 

\begin{figure}[!t]
    \centering
    \includegraphics[scale=0.9]{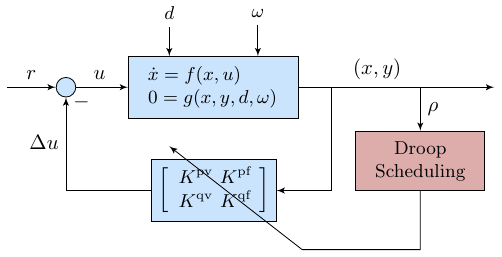}
    \caption{The online droop scheduling principle: the fast local response is obtained by introducing droop control on voltage and frequency; the gains are regularly updated (slower timescale) to meet the regulation requirements.}
    \label{fig:feedback_scheduling}
    \vspace{-0.35cm}
\end{figure}

The approach proposed in this work is illustrated in Fig.~\ref{fig:feedback_scheduling}, where an additional feedback loop is added to the droop-controlled system to update the droop slopes adaptively based on the scheduling variable $\rho$, which reflects the observable states of the system, e.g., bus voltages and DER injections. The droop gains are updated in a centralized or distributed fashion at time instants $\{k\tau_{s}\}_{k\in\mathbb{N}}$, where $\tau_{s}\in\R_{\geq 0}$ is the time required to compute and broadcast the droop gains to the individual DERs. Therefore, even though a simple proportional control law is used for the local DER control, the nonlinear structure of the system is captured by the scheduling variable $\rho$ and optimized via the additional droop scheduling loop.

\subsection{Optimal Droop Tunning via Divide and Conquer Strategy} \label{subsec:divide_and_conquer}
The droop tuning problem is solved in the spirit of gain scheduling \cite{Leith2000SurveyOG}, where the nonlinear control design task is \textit{divided} into a number of linear subproblems, each defined by an equilibrium point of \eqref{eq:sys_dae} and parametrized by a value of the scheduling variable $\rho$. Subsequently, each subproblem is \textit{conquered} separately by designing a linear controller which ensures appropriate closed-loop performance when employed with the system linearization. In this work, the linear controller design is carried out online at each timestep $t\in\{k\tau_{s}\}_{k\in\mathbb{N}}$ upon obtaining the most recent measurements $\rho_t\coloneqq (v_t,r_t,\omega_t)$, with $r_t = K_\mathrm{agg}^\mathrm{pf}$ being the reference droop gain provided by the TSO and used in \eqref{eq:pfc_provision}. Therefore, the equilibrium operating points of the plant are parametrized by the scheduling variable. 

\subsubsection{Linearization} The system dynamics in \eqref{eq:sys_dae} are approximated around a specific equilibrium point (defined by a value of $\rho$) by leveraging the approximation introduced in \eqref{eq:LinDistFlow} and setting the derivative terms to zero, as follows:
\begin{subequations} \label{eq:step1_linearization}
\begin{align}
    0 &= x - u, \\
    0 & = - v + [R,X]x + \xi + v_0(\rho), \\
    0 & = P_0(\rho) + Hx - K^\mathrm{pf}_\mathrm{agg}\Delta\omega(\rho), \label{eq:lin_p0}
\end{align}
\end{subequations}
where $\xi\coloneqq [R,X] d$ is a random variable representing uncontrollable injections. Equation~\eqref{eq:lin_p0} represents a linearization of the active power regulation requirement in \eqref{eq:pfc_provision}, where $\Delta p_\mathrm{f}$ is linearized as a function of $x$, i.e., active and reactive power injections, similarly to voltages in \eqref{eq:LinDistFlow}.
Here, $H\in\R^{2n}$ is a vector projecting DER active and reactive power injections to the PCC active power flow, and $P_0(\rho)\in\R$ is the linearization constant pertaining to the operating point $\rho$. 

\subsubsection{Optimal Droop Tuning} For an arbitrary operating point, a linear controller of the form: $u = K^\mathrm{v}\Delta v + K^\mathrm{f}\Delta\omega$, with $K^\mathrm{v}\coloneqq[K^\mathrm{pv},K^\mathrm{qv}]^\mathsf{T}$, and $K^\mathrm{f}\coloneqq[K^\mathrm{pf},K^\mathrm{qf}]^\mathsf{T}$, is designed to ensure appropriate closed-loop performance with the considered system linearization. Substituting the linear (droop) control law from \eqref{eq:DER_inputs} to \eqref{eq:step1_linearization} and rearranging, we obtain: 
\begin{align}
     v &= [R,X]K^\mathrm{v}\Delta v + [R,X]K^\mathrm{f}\Delta\omega(\rho) + \xi + v_0(\rho),\label{eq:voltage_nonlinear} \\
     0 &= P_0(\rho) + HK^\mathrm{v}\Delta v + H K^\mathrm{f}\Delta\omega(\rho) \label{eq:regulation} -K^\mathrm{pf}_\mathrm{agg}\Delta\omega(\rho).
\end{align}
The obtained relationships can now be used for the design of voltage and frequency constraint functions in the optimal droop tuning problem.
Let vector $\kappa^\mathrm{v}\in\R^{2n_c}$ collect the droop gains pertaining to voltage control, $\kappa^\mathrm{f}\in\R^{2n_c}$ represent the droop gains pertaining to frequency control, and $\kappa\coloneqq(\kappa^\mathrm{v},\kappa^\mathrm{f})$.
The droop gains $\kappa_t$ at each time step $t$ and a specific operating point $\rho_t$ are obtained by solving the following stochastic optimization problem:
\begin{subequations}\label{eq:target_opt}
\begin{align}
& \underset{\kappa_t}{\min} &&\mathbb{E}_{\xi_t} \, \Big[\,\norm{c^\mathrm{v}_t(\kappa_t^\mathrm{v})}^2_2+\norm {c_t^\mathrm{f}(\kappa^\mathrm{f}_t)}_2^2\,\Big],\\
& \,\,\textrm{s.t.}&&\mathbb{P}\Big\{h(\kappa_t^\mathrm{v},\xi_t)\leq 0 \Big\} \geq 1-\beta, \label{eq:voltreg_random_upper}\\
& && e^\mathrm{min} \leq e_t(\kappa_t^\mathrm{f}) \leq e^\mathrm{max},\label{eq:freqreg}\\
& &&\kappa_t \in \mathcal{K}_t, \label{eq:stab_cond_stochastic}
\end{align}
\end{subequations}
where the operator $\mathbb{P}\{\cdot\}$ indicates a transformation of the inequality constraint into a chance constraint, and $\mathbb{E}_{\xi_t}$ denotes expectation of a random variable $\xi_t$.
The objective function aims to minimize the voltage and frequency regulation costs, represented by the following linear maps: $c^\mathrm{v}_t(\kappa^\mathrm{v}_t):\R^{n_c}\rightarrow\R^{n_c}$, and $c^\mathrm{f}_t(\kappa^\mathrm{f}_t):\R^{n_c}\rightarrow\R^{n_c}$. The stochastic formulation is adopted for the droop gain selection problem to ensure robustness to linearization errors, measurement noises and uncertain injections of the uncontrollable devices.

The inequality constraint function $h(\kappa_t^\mathrm{v},\xi_t)$ corresponds to \eqref{eq:voltage_nonlinear} and bounds the voltage between the minimum $v^\mathrm{min}$ and the maximum $v^\mathrm{max}$. Considering the presence of the stochastic injections of uncontrolled devices, the droop gains should be scheduled in a way that voltage limits are satisfied with the prescribed probability of $1-\beta$. The relationship in \eqref{eq:voltage_nonlinear} represents a fixed point equation in $v$ and is nonlinear since $v$ appears on both sides of the equality.  Linearization of $h(\kappa_t^\mathrm{v},\xi_t)$ is proposed in the following subsection.

The frequency support provision error $e_t(\kappa_t^\mathrm{f})$ is defined using \eqref{eq:regulation}, as follows:
\begin{equation}\label{eq:provision_error}
    e_t(\kappa_t^\mathrm{f})= P_0 + HK^\mathrm{v}_{t-1}\Delta v + H K^\mathrm{f}_t\Delta\omega -K^\mathrm{pf}_\mathrm{agg}\Delta\omega,
\end{equation}
with dependences on the scheduling variable $\rho_t$ dropped for convenience. We decouple the control tasks related to frequency and voltage control by ensuring that the voltage control related gains do not appear in equations for frequency control and vice versa. To this end, the voltage droop gains from previous timesteps are used as feedforward terms to avoid crosscoupling. Therefore, the second term in \eqref{eq:provision_error} features voltage droop gains from the previous time step. A small regulation (i.e., tracking) error is allowed by $(e^\mathrm{min},e^\mathrm{max})\in\R^2$ constants in \eqref{eq:freqreg}. 
Finally, frequency and voltage (Theorem~\ref{theorem:1}) stability conditions represented by a set of feasible gains $\mathcal{K}_t$ need to be satisfied, as given in \eqref{eq:stab_cond_stochastic}.

The divide and conquer strategy therefore generates a set of proportional controllers (i.e., droop gains) by solving the optimization problem in \eqref{eq:target_opt} for each encountered operating point. In general, there are a variety of classic ways to reformulate the chance constraints \eqref{eq:voltreg_random_upper} to obtain subproblems that can be solved in real-time. These include assuming a specific functional form for the distribution (e.g., Gaussian) of $\xi$ based on the statistical information of the historical data and using constraint violation risk metrics, such as those encoding value at risk (i.e., chance constraints), Conditional Value at Risk (CVaR), distributional robustness, and support robustness. In the rest of this paper, the voltage constraints \eqref{eq:voltreg_random_upper} will be approximated by leveraging the sample average of CVaR values \cite{Rockafellar2000}. In addition, we will present an online algorithm to solve the time-varying optimal gain scheduling problem in real-time. Completely solving the problem in \eqref{eq:target_opt} at every time step might introduce excessive computational burden and is thus avoided. The online pursuit results will be adaptively applied to tune the proportional controllers. In this way, a nonlinear controller with the desired regulation properties is constructed by following the divide and conquer strategy.

\subsection{Online Primal-Dual Gradient Solution} \label{subsec:OFO}
Instead of solving the centralized nonconvex optimization problem given by \eqref{eq:target_opt} at every time step and for every encountered operating point, we employ a primal-dual gradient-based online algorithm to develop a computationally efficient feedback controller. To this end, we construct a convex surrogate of the proposed stochastic optimization problem by linearizing the constraint in \eqref{eq:voltreg_random_upper}. A similar relationship to \eqref{eq:voltage_nonlinear} was reached in \cite{KyriDroop2018}, where the nonlinearity in $\kappa^\mathrm{v}$ was tackeld using the Neuman series of a matrix. Here, we adopt a different approach, and linearize the constraint by setting $\Delta v$ on the right-hand-side of \eqref{eq:voltage_nonlinear} to be a constant value obtained from the latest measurements. For purposes of formulating the linear voltage inequality constraint, we rewrite \eqref{eq:voltage_nonlinear} as follows: $v_t(\kappa^\mathrm{v}_t,\xi_t) = v_t(\kappa^\mathrm{v}) + \xi_t$,
where $v_t(\kappa^\mathrm{v}_t)\coloneqq [R,X]K^\mathrm{v}_t \Delta v(\rho_t) + [R,X]K^\mathrm{f}_{t-1}\Delta\omega(\rho_t)+v_0(\rho_t)$ is the deterministic part of the voltage model. A feedforward term $K_{t-1}^\mathrm{f}$ is used to decouple the frequency control related gains from the voltage control task. Thus, the linear counterpart of the voltage constraint function takes the following form: $\bar{h}(v_t(\kappa^\mathrm{v}_t),\xi_t) \coloneqq [-v_t(\kappa^\mathrm{v}_t) - \xi_t + v^\mathrm{min}, v_t(\kappa^\mathrm{v}_t) + \xi_t - v^\mathrm{max}]$. For convenience, the frequency regulation inequality is written similarly, as $r(e_t(\kappa^\mathrm{f}_t)) \coloneqq [-e_t(\kappa^\mathrm{f}_t) + e^\mathrm{min}, e_t(\kappa^\mathrm{f}_t) - e^\mathrm{max}]$.

Furthermore, the chance constraints pertaining to voltage regulation can be approximated by leveraging the sample average approximation of CVaR\cite{Rockafellar2000}. The CVaR metric is a widely adopted risk measure for optimization under uncertainties \cite{Summers2015}, and the sample average approximation methods have been shown to perform well in many applications where a moderate number of the random variable samples are available~\cite{bertsimas2018robust,dall2017chance}. According to this approach, for each scenario $ s\in\Xi_t$, the following two constraints are imposed to approximate the probabilistic constraint \eqref{eq:voltreg_random_upper}:
\begin{subequations}\label{eq:ssa_cvar}
\begin{align} 
    &\frac{1}{N_s}\sum_{s=1}^{N_s}\Big[v_t(\kappa^\mathrm{v}_t) - v^{\mathrm{max}} + {\xi}_t^s + \overline{{\tau}}_t \Big]_+ - \overline{{\tau}}_t\beta \leq 0,\\
    &\frac{1}{N_s}\sum_{s=1}^{N_s}\Big[v^{\mathrm{min}} - v_t(\kappa^\mathrm{v}_t) - {\xi}_t^s + \underline{{\tau}_t} \Big]_+ - \underline{{\tau}}_t\beta \leq 0,
\end{align}
\end{subequations}
where vectors $\overline{{\tau}}_t\in\mathbb{R}^n,\underline{{\tau}}_t\in\mathbb{R}^n$ are the CVaR auxiliary variables, and $\Xi_t\coloneqq\{\xi_t^s\}_{s=1}^{N_s}$ is the set of $N_s\in\N$ samples of the random variable $\xi_t$ at time $t$. The set of constraints defined by \eqref{eq:ssa_cvar} for all $s\in\Xi_t$ will be denoted by $l_t(v_t(\kappa^\mathrm{v}_t),\tau_t)$, with $\tau_t\coloneqq(\overline{\tau}_t,\underline{\tau}_t)$. It now follows that the problem \eqref{eq:target_opt} can be approximated by a deterministic convex surrogate of the form:
\begin{subequations}\label{eq:stochastic_ds_approx}
\begin{alignat}{3}
& \underset{\kappa_t}{\min}\quad &&  C_t(\kappa_t),\\
& \,\,\textrm{s.t.} && l_t\big(v_t(\kappa_t^\mathrm{v}),\tau_t\big)\leq 0, \label{eq:voltreg_random_upper_approx}\\
& && r\big(e_t(\kappa^\mathrm{f}_t)\big) \leq 0, \label{eq:freqerg_approx} \\
& && \kappa_t \in \mathcal{K}_t,
\end{alignat}
\end{subequations}
where $C_t(\kappa_t)\coloneqq\norm{c_t^\mathrm{v}(\kappa_t^\mathrm{v})}^2_2+\norm {c_t^\mathrm{f}(\kappa^\mathrm{f}_t)}_2^2$.
To solve the above quadratic problem via a gradient approach, we consider the following regularized Lagrangian function:
\begin{equation}\label{eq:l_op}
\begin{aligned}
    &\mathcal{L}_t(\kappa_t,\tau_t,\mu_t,\lambda_t)  = C_t(\kappa_t) + \mu_t^\top l_t(v_t(\kappa_t^\mathrm{v}),\tau_t) \\
     &+ \lambda_t^\top r(e_t(\kappa^\mathrm{f}_t)) -\frac{\phi}{2}\norm{\mu_t}_2^2 -\frac{\psi}{2}\norm{\lambda_t}_2^2+ \frac{\gamma}{2}\norm{\tau_t}_2^2,
\end{aligned}
\end{equation}
where $\mu_t \in \mathbb{R}^{2n}_{\geq0}$, and $\lambda_t\in\R_{\geq0}$ are Langrange multipliers associated with constraints \eqref{eq:voltreg_random_upper_approx} and \eqref{eq:freqerg_approx}, respectively. The Tikhonov regularization terms are scaled by small constants $\psi>0$, $\gamma>0$ and $\phi>0$. The Lagrangian \eqref{eq:l_op} is strictly convex in $\kappa_t $ and $\tau_t$, and strictly concave in $\mu_t$. Therefore, gradient-based approaches can be applied to the following saddle-point problem: 
\begin{equation}\label{eq:saddle_point}
\max_{\mu_t\in\mathbb{R}^{2n}_{\geq0},\lambda_t\in\mathbb{R}_{\geq0}} \min_{\kappa_t\in\mathcal{K}_t,\tau_t\in\R^{2n}_{\geq0}} \mathcal{L}_t(\kappa_t,\tau_t,\mu_t,\lambda_t),
\end{equation}
to find an approximate solution to \eqref{eq:stochastic_ds_approx}. The difference between the optimal solutions of the original problem \eqref{eq:stochastic_ds_approx} and the regularized problem \eqref{eq:saddle_point} was characterized in \cite{koshal2011multiuser}. On the other hand, the Tikhonov regularization terms render the Lagrangian strongly convex, which enhances the convergence of \eqref{eq:saddle_point} compared to \eqref{eq:stochastic_ds_approx}.
We use the primal-dual gradient method to solve the saddle-point problem, with gradient updates given as follows:
\begin{subequations}\label{eq:primal_dual_update}
\begin{align}
\kappa_{t+1} &=\mathrm{proj}_{\mathcal{K}_t}\big[\kappa_t - \alpha\big(\nabla_{\kappa}C_t(\kappa_t) + s_t(\mu_t,\lambda_t) \big)\big],\label{eq:primal_k}\\
\tau_{t+1} &= \mathrm{proj}_{\mathbb{R}_{\geq0}^{2n}}\big[\tau_t - \alpha\big(d_t(\mu_t) + \gamma\tau_t\big) \big], \label{eq:primal_tau} \\
\mu_{t+1} & = \mathrm{proj}_{\mathbb{R}_{\geq0}^{2n}}\big[\mu_t + \alpha\big(l_t(v_t(\kappa_t^\mathrm{v}),\tau_t)- \phi\mu_t\big) \big]\label{eq:dual_mu}, \\
\lambda_{t+1} & = \mathrm{proj}_{\mathbb{R}_{\geq0}}\big[\lambda_t + \alpha\big(r(e_t(\kappa^\mathrm{f}_t)) - \psi\lambda_t\big) \big], \label{eq:dual_lambda}
\end{align}
\end{subequations}
where the operator $\mathrm{proj}_{\mathcal{X}}[x]$ projects a vector $x$ onto a set $\mathcal{X}$, and $\alpha>0$ is the update step-size. Signals $s_t(\mu_t,\lambda_t)\in\R^{2n}$ and $d_t(\mu_t)\in\R^{2n}$ contain derivatives of the constraint functions, which can be computed as follows:
\begin{align}
    s_t(\mu_t,\lambda_t) &= \nabla_{\kappa}l_t\big(v_t(\kappa_t^\mathrm{v}),\tau_t\big)^\top\mu_t + \nabla_{\kappa}r(e_t(\kappa^\mathrm{f}_t))^\top\lambda_t,\label{eq:st}  \\
    d_t(\mu_t) &= \nabla_{\tau}l_t\big(v_t(\kappa_t^\mathrm{v}),\tau_t\big)^\top\mu_t.\label{eq:dt}
\end{align}
By iteratively evaluating primal and dual variables in \eqref{eq:primal_dual_update}, the solution to the saddle point problem \eqref{eq:saddle_point} is pursued. As mentioned earlier, the introduced regularization terms create a bounded gap to the optimal solution of \eqref{eq:stochastic_ds_approx}. Convergence properties and guarantees of a similar algorithm were analyzed in our previous work~\cite{guo2022online} and have thus been omitted here.

The necessary steps for a centralized implementation of the proposed online feedback droop scheduling strategy are presented in Algorithm~\ref{algorithm:OFDS}. The algorithm can likewise be implemented in a distributed manner, in which case the local measurements are exchanged between the DERs and the update steps are performed locally. Comparison between the two approaches and a more detailed discussion is provided in \cite{Emiliano2018}. The algorithm additionally possesses plug-and-play properties as the structure of the algorithm does not change with an additional unit added, but solely \eqref{eq:primal_k} needs to be augmented with an extra pair of droop gains. Finally, the computational efficiency of the proposed algorithm lies in the fact that it is sufficient to evaluate the matrix multiplications in \eqref{eq:primal_k} instead of entirely solving the optimization problem \eqref{eq:target_opt}. Therefore, the approach does not require hardware with advanced computation capabilities, but microprocessors with capability of performing matrix multiplications are sufficient.

\algrenewcommand{\alglinenumber}[1]{\scriptsize \textbf{[S#1]}}%
\begin{algorithm}[!t]
    \caption{(Online Feedback Droop Scheduling Algorithm)}\label{algorithm:OFDS}
    \begin{algorithmic}[1]
        \State The network operator collects measurements of bus voltages, active power exchange with the upper-level grid, and the network frequency. That is, updates the value of the scheduling variable $\rho_t$, for the current time step $t$.
        \State Using the obtained measurements, dual variables $\mu_{t}$ and $\lambda_t$ can be evaluated \eqref{eq:dual_mu}-\eqref{eq:dual_lambda}.
        \State The network operator evaluates (according to \eqref{eq:st}-\eqref{eq:dt}) and distributes signals $s_t(\mu_t,\lambda_t)$ and $d_t(\mu_t)$ to DERs participating in voltage and/or frequency regulation.
        \State The DERs update their droop gain coefficients locally using \eqref{eq:primal_k} and the CVaR auxiliary variable using \eqref{eq:primal_tau}.
        \State The network operator monitors the quality of regulation by collecting further measurements and the process is repeated from [S1].
\end{algorithmic}
\end{algorithm}

\section{Results} \label{sec:res}
\begin{figure}[!t]
    \centering
    \includegraphics[scale=0.675]{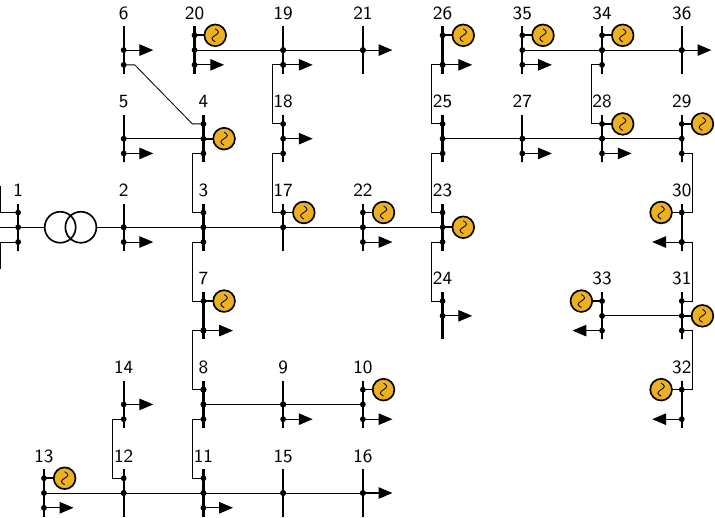}
    \caption{Single line diagram of the modified IEEE 37-bus test system.}
    \label{fig:37bus_diag}
    \vspace{-0.35cm}
\end{figure}

\begin{figure}[!b]
	\vspace{-0.35cm}
	\centering
	\begin{minipage}{0.45\textwidth}
		\centering
		\hspace{-0.4cm}
		\scalebox{1.075}{\includegraphics[]{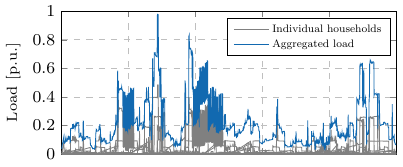}}\\
		\vspace{-0.1cm}  
	\end{minipage}
	\begin{minipage}{0.45\textwidth}    
		\centering
		\hspace{0.4cm}
		\scalebox{1.075}{\includegraphics[]{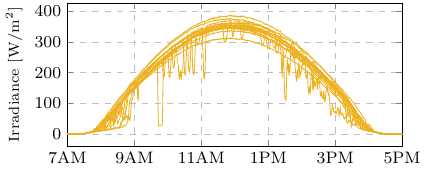}}\\
	\end{minipage}
	\caption{\label{fig:solar_load}High resolution (1-second) aggregated load demand (top) and solar irradiation (bottom) profiles over a 10-hour period.}
\end{figure}

The performance of the proposed online droop scheduling scheme is first evaluated using the IEEE 37-bus test system \cite{Schneider2018}. A balanced, single-phase equivalent of the test system is employed, with 17 PV systems and 26 load equivalents placed at different locations in the network, as illustrated in Fig.~\ref{fig:37bus_diag}. The solar irradiance profiles of 1-second resolution, measured at a site in Varennes (Qu\'ebec) and presented in \cite{GAGNEPVDATA}, are used to create the PV generation patterns. The ratings of the PV inverters are all set to \SI{200}{\kilo\voltampere}, except for the inverters at nodes 3, 15, and 16, which are \SI{350}{\kilo\voltampere}. The load demand profile is obtained from the DEDDIAG dataset \cite{Wenninger2021DEDDIAGAD}, which contains 1-second resolution consumption measurements of various appliances across 15 households. The vector of uncontrollable load injections is thus constructed by aggregating the households and rescaling them to match the base load of the 37-bus network. The resulting load demand and PV generation profiles for a 10-hour period are plotted in Fig~\ref{fig:solar_load}. The DER units are parametrized by setting the inner control loop time constants to $\tau_{p_i}=\tau_{q_i}=\SI{200}{\milli\second}$, and the reactive power setpoints to $q_{c_i}^\star=0$, for all $i\in\mathcal{C}$. The active power setpoints are determined such that the maximum power is injected considering the available solar irradiance. Considering that the inverter dynamics are taken into account and that the distribution system is modeled as a set of nonlinear DAEs \eqref{eq:sys_dae}, a DAE solver needs to be used for numerical simulations. The DAEs are solved using the nested approach~\cite{Ascher1998ComputerMF}, with the algebraic part of the model solved using the Backward-Forward Sweep method \cite{Teng2003}, and the accompanying ordinary differential equations solved using the forward Euler method. 


\subsection{Voltage Control via Droop Scheduling}
\begin{figure}[!b]
\vspace{-0.35cm}
	\centering
	\begin{minipage}{0.45\textwidth}
		\centering
 		\hspace{-0.65cm}
		\scalebox{1.075}{\includegraphics[]{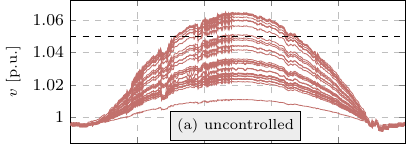}}\\
		\vspace{-0.1cm}  
	\end{minipage}
	\begin{minipage}{0.45\textwidth}
		\centering
 		\hspace{-0.65cm}
		\scalebox{1.075}{\includegraphics[]{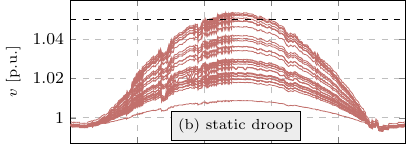}}\\
		\vspace{-0.1cm}  
	\end{minipage}
	\begin{minipage}{0.45\textwidth}    
		\centering
 		\hspace{-0.2cm}
		\scalebox{1.075}{\includegraphics[]{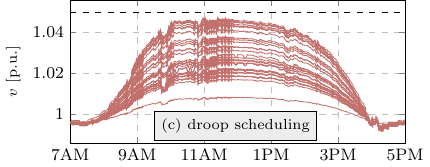}}\\
	\end{minipage}
	\caption{\label{fig:v_comp}Evolution of the network bus voltages on a clear sky day: (a) without voltage control; (b) using static droop control; (c) using online droop scheduling.}
\end{figure}

In this section, we demonstrate how the proposed droop scheduling algorithm can reliably prevent overvoltages in hours with high PV injections. While the frequency control provision is also considered, its analysis will be omitted here and thoroughly provided in the following subsection. The allowable voltage limits are set to $v_\mathrm{min}=0.95\,\mathrm{p.u.}$ and $v_\mathrm{max}=1.05\,\mathrm{p.u.}$ With the considered network setup and voltage bounds, the uncontrolled voltage response exhibits the upper limit violations at multiple buses in the time period from roughly $10$ AM to $2$ PM, as can be seen from Fig.~\ref{fig:v_comp}(a).

Other controller parameters pertaining to voltage regulation are selected as follows: the gradient step size $\alpha$ is set to $0.8$ for primal updates and $0.4$ for dual updates; regularization parameter $\phi$ is set to $3\times10^{-4}$; the chance constraint parameter $\beta=0.1$, yielding constraint satisfaction with the probability of $90\%$. The dataset $\Xi_t$ is generated at each time step $t$ by drawing $100$ samples from a Gaussian distribution with zero mean and $1.5\%$ standard deviation of the voltage magnitude $v_t$. The voltage regulation cost functions $c_t^\mathrm{v}(\kappa_t^\mathrm{v})$ weigh active power gains with $0.3$ and reactive power gains with $0.1$, thus incentivizing the use of reactive power. The controller takes measurements, computes, and updates the droop gains every $\tau_s=\SI{30}{\second}$ to take into the account limitations and delays potentially existing in the communication infrastructure.

\begin{figure}[!t]
    \centering
    \scalebox{1.075}{\includegraphics{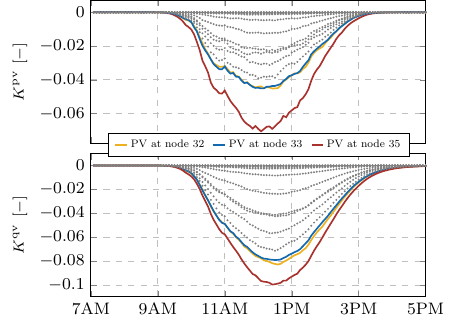}}
    \caption{Obtained droop schedule using the proposed algorithm: (i) active power - voltage gains (top); (ii) reactive power - voltage gains (bottom).}
    \label{fig:kv_droopscheduling}
    \vspace{-0.35cm}
\end{figure}
The proposed approach is first compared against a \textit{static} droop control method, where active and reactive power droops are not time-varying, i.e., they are selected offline based on network parameters and forecasts and kept constant throughout the entire control period. This approach is suggested by the IEEE 1547-2018 standard~\cite{IEEE1547} and has been considered in numerous previous works \cite{Jahangiri2013}. A common parametrization of the droop coefficients is used for the static method, where the droop coefficients are set to $0.02$ both for active and reactive power-related droop gains. Figure~\ref{fig:v_comp}(b) shows simulation results for the static droop control method. It can be seen that this control approach achieves appropriate voltage regulation, except for the interval between $11$ AM and $1$ PM. In contrast to this, the proposed droop scheduling scheme enforces voltage regulation over the entire control period as seen in Fig.~\ref{fig:v_comp}(c), leaving a small security margin in the period between $10$ AM and $2$ PM. Furthermore, the employed droop gains at different nodes are showcased in Fig.~\ref{fig:kv_droopscheduling}. Inverters at nodes $32$, $33$, and $35$, where the highest droop gains have been allocated, are emphasized. The higher magnitude of the coefficients at units located at the feeder end indicates that these units are the most impactful for voltage control. This conclusion is aligned with the results of previous research \cite{KyriDroop2018}. It can also be observed that while $K^\mathrm{pv}$ droop gains have three times higher costs, their magnitude is comparable to the magnitude of $K^\mathrm{qv}$ droop gains. This observation indicates that active power has a higher impact on voltage regulation than reactive power for the considered network. Apart from better regulation performance, the proposed droop scheduling algorithm also employs lower control efforts compared to the static method. Evaluating the objective function $C_t(\kappa_t)$ for the proposed droop scheduling method over the entire control horizon yields a value of $0.2137$, whereas the static method exhibits a higher cost of $0.3264$. Therefore, the total control effort is reduced by $65\%$.

\begin{figure}[!b]
    \vspace{-0.375cm}
	\centering
	\begin{minipage}{0.45\textwidth}
		\centering
 		\hspace{-0.65cm}
		\scalebox{1.075}{\includegraphics[]{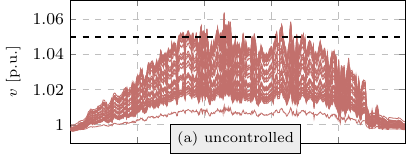}}\\
		\vspace{-0.2175cm}  
	\end{minipage}
	\begin{minipage}{0.45\textwidth}
		\centering
 		\hspace{-0.65cm}
		\scalebox{1.075}{\includegraphics[]{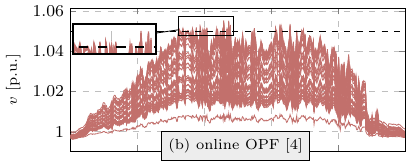}}\\
		\vspace{-0.1cm}  
	\end{minipage}
 	\begin{minipage}{0.45\textwidth}
		\centering
 		\hspace{-0.65cm}
		\scalebox{1.075}{\includegraphics[]{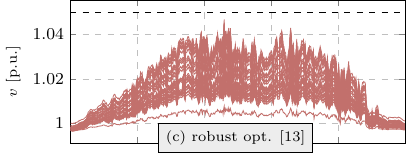}}\\
		\vspace{-0.1cm}  
	\end{minipage}
	\begin{minipage}{0.45\textwidth}    
		\centering
 		\hspace{-0.275cm}
		\scalebox{1.075}{\includegraphics[]{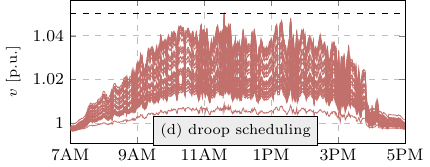}}\\
	\end{minipage}
	\caption{\label{fig:v_comp_opf}Evolution of the network bus voltages on a variable sky day: (a) without voltage control; (b) using online OPF control \cite{Emiliano2018}; (c) using adaptive droop tuning via robust optimization \cite{KyriDroop2018}; (d) using online droop scheduling.}
\end{figure}
In the following, we compare the proposed approach against a recent online OPF scheme from~\cite{Emiliano2018}, where setpoints of DERs are continuously updated to drive the inverter power outputs to AC OPF solutions, and a robust optimization-based adaptive droop tuning scheme \cite{KyriDroop2018}. For the purposes of this case study, PV irradiance data of a variable sky day from \cite{GAGNEPVDATA} is used. The obtained voltage profile when no control actions are taken is shown in Fig.~\ref{fig:v_comp_opf}(a). Even though reference \cite{Emiliano2018} suggests setpoint updates with a 1-second time step, for fair comparison and taking into account communication bottlenecks, the online OPF scheme is configured to send the setpoint updates every \SI{30}{\second}. The obtained voltage profiles when employing online OPF, robust optimization-based adaptive droop tuning, and online droop scheduling are shown in Fig.~\ref{fig:v_comp_opf}. As can be seen from the figure, the online OPF scheme struggles to contain voltages when the network is subjected to very volatile  PV injections. The lack of performance results from the fact that the setpoints are updated only every \SI{30}{\second}, and that in between the time steps, the voltage support is not actively provided. On the other hand, droop control reacts proportionally to the voltage deviation and provides voltage control at all times. Therefore, the two considered droop-based schemes capitalize on regular parameter updates and fast local control, driving the system toward optimal operation. It can furthermore be observed that the robust optimization-based control strategy employs significantly more control effort compared to the droop-scheduling approach, as the voltages are kept well below the allowable limit. Additionally, more computational effort is necessary for the solution of the robust optimization since the optimization problem is solved to convergence, whereas in the proposed droop scheduling approach, only a single primal-dual update is performed at each time step.

\subsection{Frequency Control \& Plug-and-Play Features}
\begin{figure}[!b]
    \centering
    \scalebox{1.075}{\includegraphics{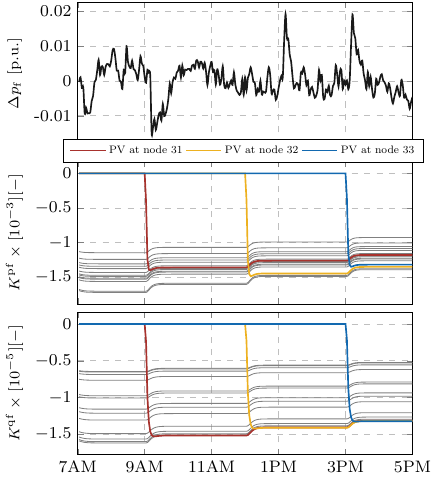}}
    \caption{Frequency control: (a) active power exchange at the interface; (b) active power - frequency droop gains; (c) reactive power - frequency droop gains.}
    \label{fig:kf_droopscheduling}
\end{figure}
The frequency control problem considered in this work is reflected in tracking the prescribed droop gain at the interface to the upstream network by a collection of DER units, rather than in optimizing the frequency response metrics. In this section, we analyze how the proposed algorithm enables DERs to pursue this goal. The droop gain to be matched by the DERs is $K_\mathrm{agg}^\mathrm{pf}=0.02$, and the 1-second resolution frequency signal is obtained from~\cite{FrequencyData}. The resulting power exchange \eqref{eq:pfc_provision} is shown in Fig.~\ref{fig:kf_droopscheduling}(a). The gradient step size parameter $\alpha$ for the frequency control droop gain update is set to $0.8$ for primal updates and $0.4$ for dual updates. The linear cost coefficients are randomly selected for each unit from interval $[0.9,1.1]$. Irradiance data of a clear sky day is again used here. In an effort to make the case study more compelling, plug-and-play features of the algorithm are demonstrated in this section. To this end, three PV units connect as ancillary service providers at different times of the day: PV at node 31 at $9$ AM, PV at node 32 at $12$ PM, PV at node 33 at $3$ PM.      

Scheduling of the frequency control-related droop gains is shown in Fig.~\ref{fig:kf_droopscheduling}. The three units which are plugged in at different times of the day are displayed with thick colored lines, while the rest are shown in gray. Allocation of the droop gains is dependent on the provision costs and sensitivity of the active power flow at the interface to active/reactive power injections at the respective unit node. Active power injection has about a hundred times higher sensitivity than reactive power for the considered network. However, reactive power is likely to be more impactful in low-voltage distribution networks. Additionally, the units newly added to the ancillary service providers' pool synchronize promptly, leading to a redistribution of the control effort among units. 

\begin{figure}[!b]
    \centering
    \scalebox{1.075}{\includegraphics{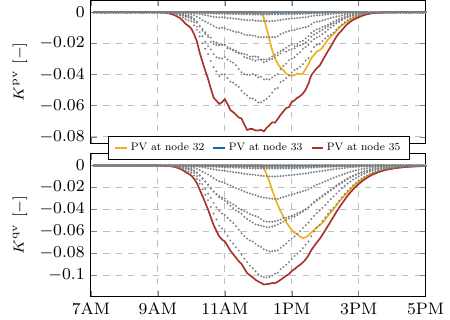}}
    \caption{Allocation of droop gains in case of connection of PV at node 31 at $9$ AM, PV at node 32 at $12$ PM, and PV at node 33 at $3$ PM.}
    \label{fig:kv_plugnplay}
\end{figure}

The voltage control performance of the proposed scheme is not notably affected when not all the units are connected from the beginning of the control period, but connect at different time instants, as considered in the present case study. That is, the voltage profile remains the same as presented in Fig.~\ref{fig:v_comp}(c). However, scheduling of the droop gains changes, with the new droop allocation presented in Fig.~\ref{fig:kv_plugnplay}. Considering that fewer voltage control providers are available in the peak PV hours, higher magnitude droop coefficients are employed compared to Fig.~\ref{fig:kv_droopscheduling}. PV at node $33$ connects at $3$ PM when PV injections have already significantly reduced and do not aid in voltage regulation. On the other hand, PV at node $32$ connects during peak hours and manages to synchronize within an hour.   

\subsection{A Large-Scale Case Study}
\begin{figure}[!b]
    \vspace{-0.375cm}
	\centering
	\begin{minipage}{0.45\textwidth}
		\centering
 		\hspace{-0.65cm}
		\scalebox{1.075}{\includegraphics[]{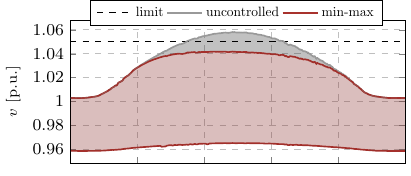}}\\
		\vspace{-0.1cm}  
	\end{minipage}
	\begin{minipage}{0.45\textwidth}
		\centering
 		\hspace{-0.65cm}
		\scalebox{1.075}{\includegraphics[]{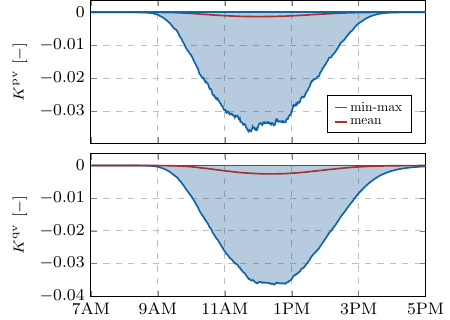}}\\
	\end{minipage}
	\caption{\label{fig:533_sim}Envelopes of the network bus voltages and DER droop gains for the 533-bus Swedish system.}
\end{figure}

To demonstrate the scalability of the proposed approach and its applicability to large-scale distribution networks, in this section, we consider a real 533-bus distribution system operated by the local DSO Kraftringen in southern Sweden. The system serves $30000$ inhabitants on an area of about 600 square kilometers and an industrial site. In order to simulate a high PV penetration scenario, the system is modified by introducing $300$ PV systems at arbitrary buses in the system. The PV injections are parameterized in the same manner as in the previous case study, whereas the original load data from \cite{Malmer2023} is preserved. The simulation results are reported in Fig.~\ref{fig:533_sim}, where the voltage magnitude and droop gain evolution over a 10-hour period are presented. Due to the large number of buses in the system, envelopes are used instead of individual trajectories. As can be seen from the figure, the droop scheduling approach successfully contains the voltages with relatively low control effort, as demonstrated by low mean values of the employed droop gains. Furthermore, it is important to emphasize that the computational time required for the primal and dual updates remains below \SI{1}{\milli\second} at each time step.

\section{Conclusion} \label{sec:concl}
This paper proposes a scalable control framework for local voltage and frequency control provision by DERs in distribution grids. The proposed approach relies on three fundamental concepts: (i) droop control - a strategy ubiquitous to power system regulation problems; (ii) feedback - a vital tool of control theory; (iii) distributed (online) optimization - a methodology critical for achieving computational efficiency and enabling scalability. With the aid of the tools mentioned above, we first equipped each DER with a generalized droop matrix to enable fast local control. Subsequently, we designed an online scheduling algorithm to update the droop gain matrices periodically and govern the system towards a desired operating point. The numerical simulations demonstrate that the proposed approach is superior to the static droop control method for voltage regulation, as it improves the regulation quality using less control effort. Furthermore, in contrast to the recently proposed online OPF schemes, our approach behaves well under highly variable PV injections. In contrast to robust optimization-based droop tuning, our approach employs less control effort and is more computationally efficient. Plug-and-play capabilities and tracking of the frequency droop gain provided by the TSO have also been demonstrated. The main drawbacks of the algorithm lie in the requirement for a network model and potential difficulties in step-size tuning. 

Finally, the stability properties of droop-controlled distribution grids were also analyzed in this paper. A Lyapunov function was constructed to identify the stability conditions for the droop gains of individual DERs. Three main conclusions have been drawn: (i) droop gains pertaining to active and reactive power are tightly interdependent; (ii) the network topology and parameters play an essential role; (iii) the obtained conditions are decentralized, i.e., there are no interdependencies between droop gains of different DERs.

\appendix
\setcounter{secnumdepth}{0}
\begin{proof}[Proof of Theorem 1]
Consider the following Lyapunov function candidate
\begin{equation} \label{eq:thm1_V}
    V(\Delta x) = \frac{1}{2}\Delta x^\top G \Delta x.
\end{equation}
Clearly, $V(\Delta x)$ is positive definite by Proposition \ref{prop:R_X_PDF}. A sufficient condition for the system \eqref{eq:closed_loop_ssm} to be stable in a Lyapunov sense is that the following condition of the derivative of the Lyapunov function \eqref{eq:thm1_V} holds:
\begin{equation} \label{eq:Lyapunov_condition_1}
\begin{aligned}
    \ddt V(\Delta x) & = \left(\nabla_x V(\Delta x)\right)^\top \Delta\dot{x}\\
    & = \left(G\Delta x\right)^\top T^{-1} \left(K_\textrm{v}G - I\right)\Delta x\\
    & = \Delta x^\top G^\top T^{-1}K_\textrm{v}G\Delta x - \Delta x^\top G T^{-1} \Delta x < 0.
\end{aligned}
\end{equation}
To establish the condition of negative derivative of the Lyapunov function, we replace the second term of the last equality by $\Delta x^\top G T^{-1}G^{-1} G \Delta x$ due to the non-singularity of $G$. The condition in \eqref{eq:Lyapunov_condition_1} boils down to
\begin{equation} \label{eq:Lyapunov_condition_2}
\begin{aligned}
    \ddt V(\Delta x) & = \Delta x^\top G^\top T^{-1}\left(K_\textrm{v} - G^{-1}\right)G\Delta x < 0,
\end{aligned}
\end{equation}
which leads to the following negative definiteness condition:
\begin{equation} \label{eq:Lyapunov_condition_3}
    T^{-1}\left(K_\textrm{v} - G^{-1} \right)\prec 0.
\end{equation}
Note that $G$ and $T$ are non-singular and the inverse of the multiplication $(GT)^{-1}$ exists. The inequality 
\begin{equation}\label{eq:Lyapunov_condition_4}
T^{-1}K_\textrm{v} - \lambda_{\textrm{min}}((GT)^{-1}) I \prec 0 
\end{equation} 
implies the condition \eqref{eq:Lyapunov_condition_3}. Expanding condition 
\eqref{eq:Lyapunov_condition_4} yields
\begin{equation}\label{eq:bmatrix_full}
\begin{bmatrix}
    T_p^{-1}K^\textrm{pv} - \gamma I &  T_p^{-1}K^\textrm{pv}\\
    T_q^{-1}K^\textrm{qv} & T_q^{-1}K^\textrm{qv} - \gamma I
\end{bmatrix} \prec 0.
\end{equation}
where $\gamma = \lambda_{\textrm{min}}((GT)^{-1})$. The block matrix in \eqref{eq:bmatrix_full} is asymmetric and only the symmetric part defines the negative definiteness, more precisely
\begin{equation*}
    \begin{bmatrix}T_p^{-1}K^\textrm{pv} - \gamma I &  (T_p^{-1}K^\textrm{pv} + T_q^{-1}K^\textrm{qv})/2 \\
    (T_p^{-1}K^\textrm{pv} + T_q^{-1}K^\textrm{qv})/2 & T_q^{-1}K^\textrm{qv} - \gamma I \end{bmatrix}\prec 0.
\end{equation*}
\noindent Supposing $T_p^{-1}K^\textrm{pv} - \gamma I \prec 0$ and leveraging the Schur Complement, we obtain
\begin{equation}\label{eq:thm1_schur}
\begin{aligned}
    &\left(T_q^{-1}K^\textrm{qv} - \gamma I\right)
     - \left(\frac{T_p^{-1}K^\textrm{pv} + T_q^{-1}K^\textrm{qv}}{2}\right)^\top \\
    &~~~~\times \left(T_p^{-1}K^\textrm{pv} - \gamma I\right)^{-1}   \left(\frac{T_p^{-1}K^\textrm{pv} + T_q^{-1}K^\textrm{qv}}{2}\right) \prec 0.
\end{aligned}
\end{equation}
Since $T_p^{-1}K^\textrm{pv}$ and $T_q^{-1}K^\textrm{qv}$ are diagonal, we expand \eqref{eq:thm1_schur} element-wise which results in
\begin{equation}\label{eq:thm1_elements_cond}
\begin{aligned}
    &\left(\tau_{p_i}^{-1}k^{\textrm{pv}}_i - \tau_{q_i}^{-1}k^{\textrm{qv}}_i\right)^2 \\
    &~~~~~~~~~ + 4\gamma \left(\tau_{p_i}^{-1}k^{\textrm{pv}}_i - \tau_{q_i}^{-1}k^{\textrm{qv}}_i\right) - 4\gamma^2 < 0, \forall i \in \mathcal{N}.
\end{aligned}
\end{equation}
Leveraging another statement of the Schur complement by assuming $T^{-1}_qK^\textrm{qv} - \gamma I \prec 0$, we arrive at the same condition as given in \eqref{eq:thm1_elements_cond}. In addition, since $G$ is non-singular, $\dot{V}(\Delta x) = 0$ if and only if $\Delta x = 0$. As a result, either $\tau_{p_i}^{-1}k^\textrm{pv}_i - \gamma < 0$ or $\tau_{q_i}^{-1}k^\textrm{qv}_i - \gamma < 0$ holds for all $i\in\mathcal{N}$ and together with condition \eqref{eq:thm1_elements_cond} implying the negative definiteness of $T^{-1}(K_\textrm{v} - G^{-1})$, which leads to $\dot{V}(\Delta x) < 0$ for any $\Delta x \neq 0$ and $\dot{V}(0) = 0$. This concludes the proof.
\end{proof}
 
\bibliographystyle{IEEEtran}
\bibliography{bibliography}

\end{document}